\documentclass{egpubl}
\usepackage{egsr2025}
 
\ConferencePaper        % uncomment for (final) Conference Paper
\WsPaper           % uncomment for final version of EG Workshop contribution
\CGFccby
\usepackage[T1]{fontenc}
\usepackage{dfadobe}  

\usepackage{cite}  % comment out for biblatex with backend=biber
\BibtexOrBiblatex
\electronicVersion
\PrintedOrElectronic
\ifpdf \usepackage[pdftex]{graphicx} \pdfcompresslevel=9
\else \usepackage[dvips]{graphicx} \fi

\usepackage{egweblnk} 
\title[Neural Path Guiding with Distribution Factorization]%
      {Neural Path Guiding with Distribution Factorization}

\author[Figueiredo et al.]
{\parbox{\textwidth}{\centering Pedro Figueiredo\orcid{0000-0002-3807-1512},
        Qihao He\orcid{0009-0000-3011-3135} and 
        Nima Khademi Kalantari\orcid{0000-0002-2588-9219}
        }
        \\
{\parbox{\textwidth}{\centering Texas A\&M University
       }
}
}
\newcommand{\vect}[1] {\mathbf{#1}}
\newcommand{\surface}{\vect{x}}
\newcommand{\diro}{\omega_o}
\newcommand{\diri}{\omega}
\newcommand{\netparam}{\Theta}
\newcommand{\netparamf}{{\vect{w}_1}}
\newcommand{\netparams}{{\vect{w}_2}}
\newcommand{\netparamg}{{\vect{w}}}
\newcommand{\nrcparam}{{\Phi}}
\newcommand{\bsdfsym}{{\rho}}

\usepackage{booktabs} % For formal tables
\usepackage{xcolor}
\usepackage{mathtools,colortbl}
\usepackage{appendix}
\usepackage{bm}

\newcommand{\rev}[1]{\textcolor{black}{#1}}

\begin{document}

\maketitle
%-------------------------------------------------------------------------
\begin{abstract}
   In this paper, we present a neural path guiding method to aid with Monte Carlo (MC) integration in rendering. Existing neural methods utilize distribution representations that are either fast or expressive, but not both. We propose a simple, but effective, representation that is sufficiently expressive and reasonably fast. Specifically, we break down the 2D distribution over the directional domain into two 1D probability distribution functions (PDF). We propose to model each 1D PDF using a neural network that estimates the distribution at a set of discrete coordinates. The PDF at an arbitrary location can then be evaluated and sampled through interpolation. To train the network, we maximize the similarity of the learned and target distributions. To reduce the variance of the gradient during optimizations and estimate the normalization factor, we propose to cache the incoming radiance using an additional network. Through extensive experiments, we demonstrate that our approach is better than the existing methods, particularly in challenging scenes with complex light transport.
\printccsdesc   
\end{abstract}  

\section{Introduction}
\label{sec:intro}

Accurate simulation of light transport necessitates computing a complex multidimensional integral. Monte Carlo (MC) rendering, a common approach for physically-based light transport simulation, estimates this integral by averaging samples of the integrand. While this approach can produce compelling effects, the variance of the estimator decreases linearly with the number of samples. Therefore, accurate estimation of the integral requires evaluating a large number of expensive samples (paths), making the approach slow.

An effective way of reducing the variance of MC rendering is through importance sampling. Specifically, the goal of importance sampling is to sample the paths with higher contribution to the integral more frequently. A large number of methods~\cite{Guo_2018,Vorba_TOG_2014,Muller_CGF_2017} focus on devising strategies to approximate sampling distributions, typically in an online fashion, that are proportional to the integrand. Particularly, the \emph{path guiding} approaches adopt a local view and perform the guided sampling at every intersection point along a path. These methods use efficient data structures, such as k-d tree~\cite{Dodik_CGF_2022},  octree~\cite{Bus_JCGT_2017}, and quadtree~\cite{Muller_CGF_2017}, as well as parametric models like Gaussian mixture model (GMM)~\cite{Herholz_CGF_2016,Vorba_TOG_2014}, and von Mises-Fisher (vMF) mixture models~\cite{Ruppert_TOG_2020}, to estimate the distributions in a data-driven manner. Most of these methods learn a marginalized distribution within each subdivision, thus failing to capture the spatial variations in each subarea.

More recently, a few techniques~\cite{Muller_TOG_2019, Dong_2023, Huang_TOG_2024} overcome the limitations in explicit representations by utilizing neural networks. The network in these methods takes a continuous coordinate (spatial as well as directional in case of product sampling) as the input, and estimates the distribution at that particular location. The key difference between these methods is in how they model the probability distribution function (PDF). Specifically, M\"{u}ller et al.~\shortcite{Muller_TOG_2019} use normalizing flows~\cite{Dinh_2017}, an expressive model with the ability to represent detailed distributions at the cost of being slower and more difficult to optimize. Dong et al.\cite{Dong_2023} and Huang et al.\cite{Huang_TOG_2024}, on the other hand, use mixture models that are relatively easy to optimize. \rev{In practice, these models are used with only a few components to ensure efficiency. Unfortunately, such models are not able to handle challenging light transport scenarios.}

In this paper, we propose a simple representation that is sufficiently expressive and reasonably fast. Specifically, we represent the joint PDF over the 2D directional domain in terms of the product of two 1D distributions: a marginal and a conditional PDF. We model each distribution using a separate neural network that estimates the PDF at uniformly spaced discrete locations. The PDF at an arbitrary coordinate is obtained by interpolating the estimated values. We explore two simple interpolation strategies and present approaches to evaluate and sample from our learned representation.

We train our networks by minimizing the loss between the predicted and target distributions. Since the target distribution is unknown, existing methods~\cite{Dong_2023,Muller_TOG_2019} use its MC samples without normalization during optimization. Because of this, the gradients are highly noisy, making the training unstable. To address this issue, we identify the incoming radiance as the source of the variance, and propose to cache it using an additional network which is in turn used to estimate a better target distribution. We demonstrate that the combination of radiance caching and our simple PDF representation is highly effective and outperforms the existing techniques, particularly in challenging scenes.
\vspace{-0.05in}
\section{Related Work}
\label{sec:relatedwork}

\paragraph*{Global Path Guiding} These approaches attempt to sample a full path (from camera to light sources) at once according to the distribution of the integrand. This is typically done in the primary sample space (PSS), which provides an elegant way of parameterizing a path, while ensuring each vertex lies on the surfaces. Various approaches propose to do this using k-d trees~\cite{Guo_2018} or neural networks~\cite{Zheng_CGF_2019, Muller_TOG_2019}. However, these approaches suffer from the curse of dimensionality and their performance degrade with longer paths~\cite{Muller_TOG_2019}. Specialized guiding methods have been developed to efficiently deal with glossy interactions, offering improved performance in specific scenarios~\cite{Li_SigAsia_2022, Fan_TOG_2023}, which is not the primary focus of our framework.

\paragraph*{Local Path Guiding} An alternative method is to perform guiding at every intersection point along the path. Early approaches represent the guiding distribution using a 2D histogram and store it through photon mapping~\cite{Jensen_EGSR_1995} or a 5D tree~\cite{Lafortune_EGSR_1995}. Follow up work refine this strategy through a combination of BRDF and photon map before sampling~\cite{Steinhurst_SIRT_2006} and improvements in caustics sampling~\cite{Budge_CGF_2008}. We also use a discretized representation, but factorize it to two 1D distributions and estimate them using neural networks. Moreover, we present evaluation and sampling using linear interpolation, in addition to the commonly-used used nearest neighbor strategy. 

A variety of other representations including hemispherical particle footprint~\cite{Hey_2002}, cosine lobes~\cite{Bashford_CGF_2012}, B-splines~\cite{Pegoraro_CGF_2008}, binary and quad tree~\cite{Muller_CGF_2017, Vorba_2019, Rath_TOG_2020, Zhu_TOG_2021}, double trees~\cite{Bus_JCGT_2017}, and discretized hemispheres using reinforcement learning~\cite{Dahm_2017} have also been proposed. Moreover, several methods use parametric models, such as Gaussian mixture~\cite{Vorba_TOG_2014,Herholz_CGF_2016, Dodik_CGF_2022, Vorba_TOG_2016} and von Mises-Fisher (vMF) mixture models~\cite{Herholz_TOG_2019,Ruppert_TOG_2020}. While most of these approaches fail to capture the spatial variations due to their limited spatial partitioning, a couple of methods address this issue~\cite{Ruppert_TOG_2020, Dodik_CGF_2022}. Ruppert et al.~\cite{Ruppert_TOG_2020} introduce a parallax compensation strategy that readjusts vMF lobes to point to a virtual light source. Unfortunately, they do not account for occlusion in such adjustments. Dodik et al.~\cite{Dodik_CGF_2022} propose a higher-dimensional representation that can perform product sampling, but this comes at the cost of increased computational cost.

More recently, path guiding methods utilizing neural networks have emerged. Instead of using complex data structures, a few techniques~\cite{Muller_TOG_2019, litalien2024, Dong_2023, Huang_TOG_2024} use neural networks to estimate the distributions in an online manner. Particularly, a couple of methods~\cite{Muller_TOG_2019, litalien2024} use normalizing flows~\cite{Dinh_2017} to map a known distribution (e.g., uniform) to the target distribution. Unfortunately, evaluating and sampling the PDF from normalizing flows is computationally expensive. A couple of methods attempt to address this problem by using vMF~\cite{Dong_2023} and normalized anisotropic spherical Gaussian (NASG)~\cite{Huang_TOG_2024} mixtures. The main problem with these two approaches is that the mixture model representations have limited expressiveness, and thus struggle in scenes with complex light transport. In contrast to all the discussed online methods, a few approaches propose to perform path guiding using convolutional neural networks (CNN)~\cite{Bako_CGF_2019, Huo_TOG_2020, Zhu_TOG_2021, Zhu_TOG_2021_B} in an offline manner. These methods train a CNN on a large number of scenes and use the trained network to adaptively refine the distribution~\cite{Huo_TOG_2020} or denoise/improve the guiding maps~\cite{Bako_CGF_2019, Zhu_TOG_2021, Zhu_TOG_2021_B} during rendering. In addition to having a large computational cost associated with the CNNs, some of these techniques~\cite{Bako_CGF_2019, Huo_TOG_2020} only allow guiding at the first intersection.

\paragraph*{Radiance Caching} Since the seminal work by Ward et al.~\shortcite{Ward_1988}, a plethora of methods aiming for both offline~\cite{Gassenbauer_CGF_2009, Dubouchet_2017, Marco_TOG_2018, Zhao_2019} and real-time~\cite{Binder_2018, Pantaleoni_2020,Scherzer_CGF_2012, Rehfeld_2014,Silvennoinen_TOG_2017,Vardis_JCGT_2014} rendering have been proposed. The key challenge in radiance caching is placement of the cache records and interpolation, particularly in presence of glossy surfaces. The recent neural radiance caching (NRC)~\cite{Muller_TOG_2021} tackles these challenges by caching the reflected radiance along an outgoing direction at each surface point using a highly efficient implementation of a multilayer perceptron (MLP). We use this approach to cache the incoming radiance, but in the context of path guiding, and for reducing the variance of the gradients and estimating the normalization factor.

\paragraph*{\rev{Histogram Prediction}} \rev{
A couple of methods use discretized histograms to model PDFs in closely related problems. For example, Zhu et al.~\cite{zhu2021neural} learn two-dimensional discrete importance weights to sample complex luminaires from different perspectives. Xu et al.~\cite{xu2023neusample} extend this idea by decomposing the histogram into weighted combinations of shared-basis histograms, allowing a more accurate representation of diverse PDF lobes in the context of neural materials. In both cases, optimization is performed as a preprocessing step, allowing ample time and data samples to effectively learn complex target PDFs. However, for online path guiding, it is critical to balance expressiveness with training and inference costs. This motivates our efficient formulation, which represents the joint 2D PDF as a product of two 1D distributions.}

\begin{figure}[t]
\centering
\includegraphics[width=1.0\linewidth, scale=1.0, angle=-0]{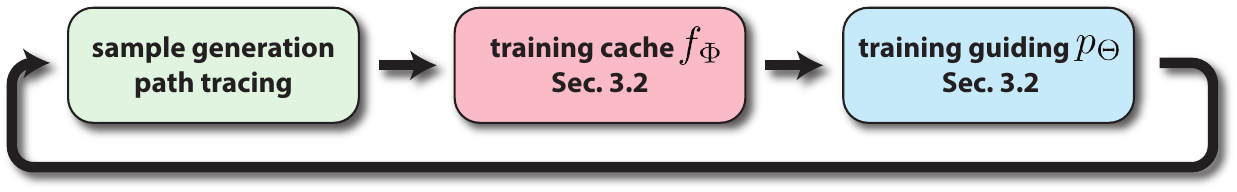}
\vspace{-0.2in}
\caption{Method Overview. During path tracing, we generate samples using our guiding distribution ${p}_\netparam$ to increase the number of paths that reach light sources. These paths are then used to train our radiance caching $f_\nrcparam$. We leverage the cached radiance as a smoother objective to improve $p_\netparam$, which in turn is used on the next sample generation.}
\vspace{-0.25in}
\label{fig:overview}
\end{figure}
\vspace{-0.05in}
\section{Method}
\label{sec:method}

We begin by formulating the light transport in a local manner using the rendering equation~\cite{Kajiya_CG_1986}. Based on this formulation, the outgoing radiance $L_o$ at a surface point $\vect{x}$ in direction $\omega_o$ can be written in terms the emitted $L_e$ and incident $L_i$ radiance as follows:

\vspace{-0.15in}
\begin{equation}
\label{eq:rendering_eq}
    L_o(\surface, \diro) = L_e (\surface, \diro) + \int_\Omega \bsdfsym(\surface, \diro, \diri) L_i(\surface, \diri) |\cos(\theta)| \ \text{d}\diri,
\end{equation}
\vspace{-0.15in}

\noindent where $\bsdfsym$ is the bidirectional scattering distribution function (BSDF), $\Omega$ is the unit sphere on $\vect{x}$ containing all the incoming directions $\omega_i$, and $\theta$ is the angle between the incoming direction and surface normal at $\vect{x}$. The key challenge here is estimating reflected radiance by computing the integral term which is typically done through MC integration as follows:

\vspace{-0.15in}
\begin{equation}
\label{eq:MC_integration}
    \langle L_r(\surface, \diro) \rangle = \frac{1}{N} \sum_{i = 1}^{N} \frac{ \bsdfsym(\surface, \diro, \diri_i) L_i(\surface, \diri_i) |\cos(\theta_i)|}{\hat{p}(\diri_i | \surface, \diro)}, 
\end{equation}
\vspace{-0.15in}

\noindent where $\hat{p}(\diri_i | \surface, \diro)$ is the distribution from which the sampled incoming directions $\omega_i$ are drawn from. To lower the variance, this distribution should be proportional to the integrand, i.e., $\hat{p}(\diri | \surface, \diro) \propto \bsdfsym(\surface, \diro, \diri) L_i(\surface, \diri) |\cos(\theta)|$.

Our goal is to model the guiding distribution using a neural network, $\hat{p}_\netparam(\diri | \surface, \diro)$ where $\netparam$ represents the network parameters, and train it in a data-driven manner. In the following sections, we explain our distribution model and optimization strategies. The overview of our entire pipeline is illustrated in Fig.~\ref{fig:overview}.

\vspace{-0.05in}
\subsection{Distribution Factorization}
\label{ssec:dist_model}

The success of path guiding methods depends in part on the choice of PDF representation. For such a representation to be effective, it should possess two main properties. 1) It should be sufficiently expressive to match the shape of the integrand (target distribution), even in complex settings. 2) Generating samples $\diri_i$ from the distribution and evaluating their PDF $\hat{p}_\netparam(\diri_i | \surface, \diro)$, which are both necessary for MC integration (Eq.~\ref{eq:MC_integration}), should be fast.

\begin{figure}[t]
\centering
\includegraphics[width=1.0\linewidth, scale=1.0, angle=-0]{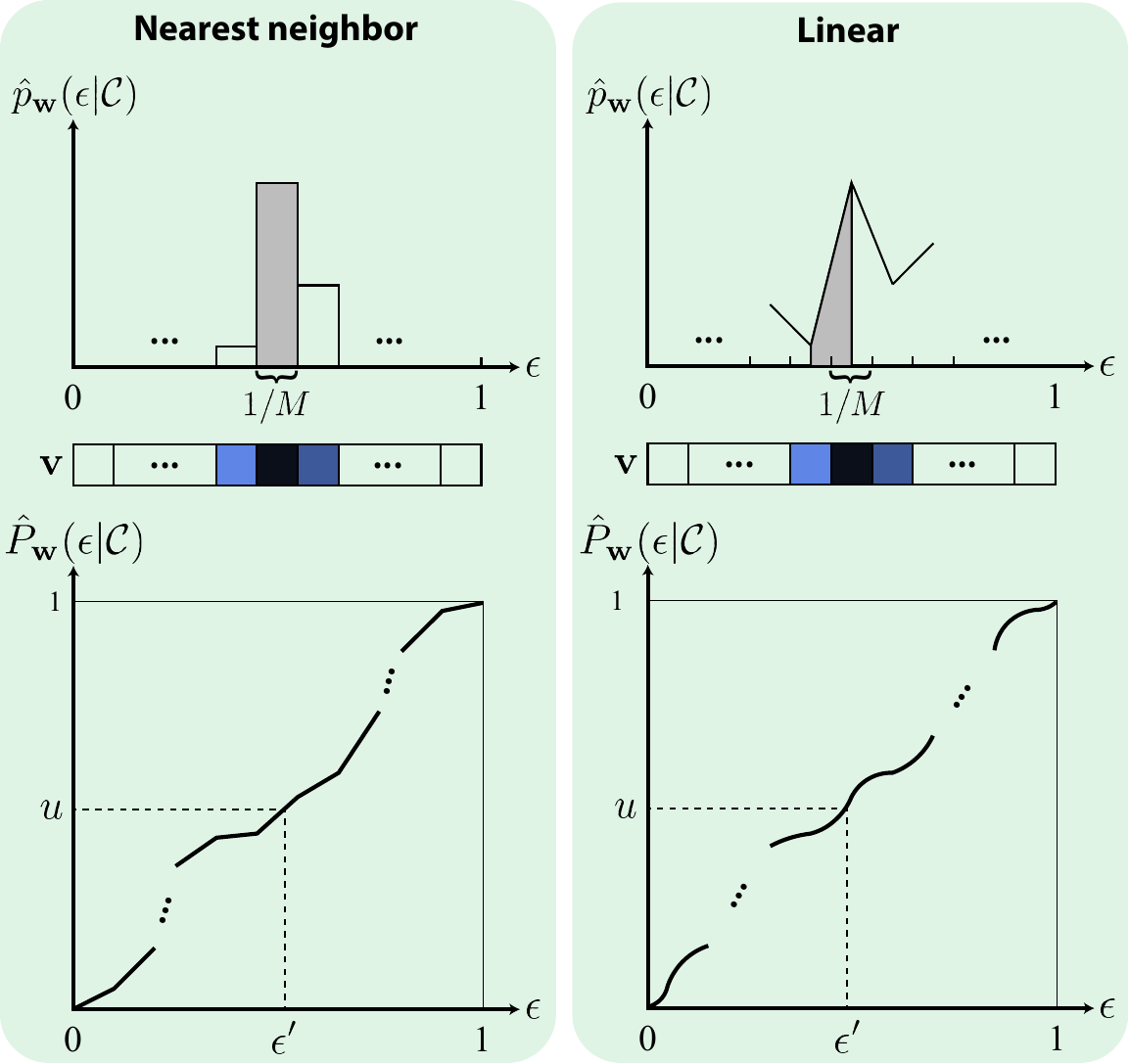}
\vspace{-0.2in}
\caption{We demonstrate the PDF evaluation and sampling process for nearest neighbor and linear interpolation. To model the 1D distributions (marginal and conditional in Eq.~\ref{eq:product_rule}) our network first predicts a vector $\vect{v}$ containing estimates of the PDF at discrete locations. To obtain the PDF at an arbitrary location, we either use the PDF estimate at the closest sample (top-left), or linearly interpolate between the two closest samples (top-right). Sampling is done by evaluating the inverse CDF at a randomly generated value $u$ with a uniform distribution. Note that the CDF for nearest neighbor interpolation is piecewise linear, while it is piecewise quadratic for linear interpolation.}
\vspace{-0.2in}
\label{fig:interpolation}
\end{figure}

To this end, we propose to model the distribution explicitly, by estimating the PDF at a set of discrete locations and obtaining the continous PDF through interpolation. To ensure the discretization is done uniformly over the directional domain, we define the PDF in uniform square space with coordinates $\epsilon_1$ and $\epsilon_2$ ranging between 0 to 1. These coordinates can be easily converted to spherical domain through $\phi = 2\pi \epsilon_1$ and $\theta = \cos^{-1}(1 - 2\epsilon_2)$. The PDF defined over the uniform square space $\hat{p}(\epsilon_1, \epsilon_2 | \surface, \diro)$ can be converted to distribution over $\diri$, $\hat{p}(\diri | \surface, \diro)$, by taking the Jacobian of the transformation into account.

Estimating the PDF over a grid of size $M_1\times M_2$ (discretization resolution) using a neural network could potentially be computationally expensive. To avoid this problem, we use the product rule to represent the joint PDF over the uniform square domain as the multiplication of a marginal and a conditional PDF as follows:

\vspace{-0.15in}
\begin{equation}
\label{eq:product_rule}
    \hat{p}_\netparam(\epsilon_1, \epsilon_2 | \surface, \diro) = \hat{p}_\netparamf(\epsilon_1 | \surface, \diro) \ \hat{p}_\netparams(\epsilon_2 | \epsilon_1, \surface, \diro).
\end{equation}
\vspace{-0.15in}

Through this representation, estimating the joint PDF boils down to predicting two 1D distributions. We discretize the two domains, $\epsilon_1$ and $\epsilon_2$, into $M_1$ and $M_2$ discrete locations, respectively, and use two independent networks to estimate the PDF at those locations. Specifically, one network $f_{\netparamf}$ models $\hat{p}_\netparamf(\epsilon_1 | \surface, \diro)$ by taking $\surface$ and $\diro$ as the input and estimating an $M_1$ dimensional vector $\vect{v}_1$, while the other network $f_\netparams$ approximates $\hat{p}_\netparams(\epsilon_2 | \epsilon_1, \surface, \diro)$ by taking $\epsilon_1$, in addition to $\surface$ and $\diro$, and predicting an $M_2$ dimensional vector $\vect{v}_2$. The PDF at an arbitrary location can then be obtained by interpolating the estimated PDFs at discrete locations.

Care should be taken when estimating the PDF vectors to ensure the interpolated continuous functions are valid distributions and integrate to one. Below, we discuss our network design to estimate valid PDFs for two interpolation strategies: nearest neighbor and linear. We explain each case in a general setting since the two networks are similar and only differ in their input conditions. In Sec.~\ref{sec:results}, we compare the performance of these two methods.

\paragraph*{Nearest Neighbor} In this case, as shown in Fig.~\ref{fig:interpolation} (top-left), the domain is divided into a set of $M$ bins and the PDF inside each bin is obtained from the corresponding element of the estimated vector $\vect{v}$ as follows:

\vspace{-0.25in}
\begin{equation}
\label{eq:nearest_neighbor}
    \hat{p}_\netparamg(\epsilon|\mathcal{C}) = \vect{v}[\lfloor\epsilon M \rfloor].
\end{equation}
\vspace{-0.15in}

\begin{figure}[t]
\centering
\includegraphics[width=1.0\linewidth, scale=1.0, angle=-0]{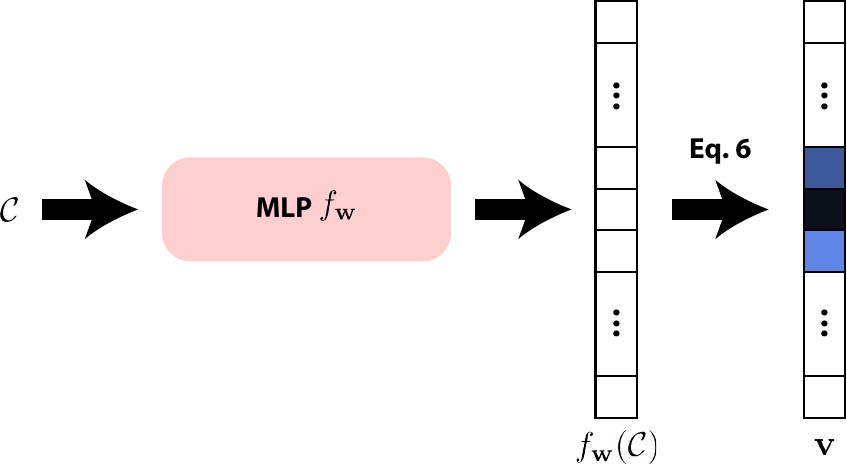}
\vspace{-0.2in}
\caption{Our multilayer perceptron (MLP) takes the condition $\mathcal{C}$ as the input and estimates an $M$ dimensional vector. We then apply softmax function to this vector and multiply each element by $M$ to obtain a vector containing the PDF estimate at discrete locations. We use this network to model the marginal and conditional distributions in Eq.~\ref{eq:product_rule}. Note that the we use $\surface$ and $\diro$ as the condition when modeling the marginal distribution, but for the conditional one, we additionally pass the first dimension $\epsilon_1$ to the network. We use two separate networks to model the two distributions, but we follow the process illustrated in this figure in both cases. }
\vspace{-0.25in}
\label{fig:network}
\end{figure}

Here, we drop the subscripts since the process for the two PDFs are similar. Moreover, $\mathcal{C}$ refers to the condition which is different for the two PDF terms in Eq.~\ref{eq:product_rule}. To have a valid PDF, the cumulative distribution function (CDF) at one ($\hat{P}_\netparamg(1)$) should be equal to one, i.e., the PDF should integrate to one. We can formally write this as:

\vspace{-0.15in}
\begin{equation}
\label{eq:nearest_integral}
    \hat{P}_\netparamg(1) = \int_0^1 \hat{p}_\netparamg(\epsilon|\mathcal{C}) \ \text{d}\epsilon = \sum_{i = 0}^{M-1} \vect{v}[i]\frac{1}{M} = 1,
\end{equation}
\vspace{-0.15in}

\noindent where the summation calculates the integral by computing the area underneath each bar, one of which shaded in gray in Fig.~\ref{fig:interpolation} (top-left). For this equation to hold, the elements of the output vector should sum up to $M$, i.e., $\sum_{i = 0}^{M-1} \vect{v}[i] = M$. To achieve this, we obtain $\vect{v}$ by applying softmax to the output of the network and multiplying each element by M as follows (see Fig.~\ref{fig:network}):

\vspace{-0.15in}
\begin{equation}
\label{eq:normalization}
    \vect{v} = M\ \sigma\big(f_\netparamg(\mathcal{C})\big),
\end{equation}
\vspace{-0.15in}

\noindent where $\sigma$ is the standard softmax operator. We can thus evaluate the PDF at an arbitrary location through combination of Eqs.~\ref{eq:nearest_neighbor}~and~\ref{eq:normalization}.

To sample from the learned distribution $\hat{p}(\epsilon | \mathcal{C})$, we use the inverse transform sampling technique. As shown in Fig.~\ref{fig:interpolation} (bottom-left), we evaluate the inverse CDF at a randomly generated number with uniform distribution $u$, i.e. $\epsilon^\prime = \hat{P}_\netparamg^{-1}(u | \mathcal{C})$. The exact formulation of the inverse CDF is provided in the supplementary material.

\paragraph*{Linear} Here, the PDF at an arbitrary location is obtained by linearly interpolating the estimated PDF at the two nearest discrete coordinates as follows (see Fig.~\ref{fig:interpolation} top-right): 

\vspace{-0.15in}
\begin{equation}
\label{eq:linear}
    \hat{p}_\netparamg(\epsilon|\mathcal{C}) = (1 - \alpha) \ \vect{v}[\lfloor m \rfloor] + \alpha \ \vect{v}[\lceil m \rceil],
\end{equation}
\vspace{-0.15in}

\vspace{-0.1in}
\noindent where
\vspace{-0.1in}

\vspace{-0.15in}
\begin{equation}
    m = \epsilon M - 0.5, \quad \text{and} \quad \alpha = m - \lfloor m \rfloor.
\end{equation}
\vspace{-0.15in}

Note that the subtraction of 0.5 is made to ensure the elements of the vector $\vect{v}$ provide PDF values at the center of each bin. Again, to ensure $\hat{p}_\netparamg$ is a valid PDF, it should integrate to one, i.e.,

\vspace{-0.15in}
\begin{equation}
\label{eq:normalization_linear}
    \hat{P}_\netparamg(1) = \int_0^1 \hat{p}_\netparamg(\epsilon|\mathcal{C}) \ \text{d}\epsilon = \sum_{i = 0}^{M - 1}\frac{\vect{v}[i] + \vect{v}[i+1]}{2}\frac{1}{M} = 1.    
\end{equation}
\vspace{-0.15in}

In this case, the integral is calculated by summation each trapezoid's area, one of which shaded in gray in Fig.~\ref{fig:interpolation} (top-right).

Note that the boundaries ($\epsilon < \frac{1}{2M}$ and $\epsilon > 1 - \frac{1}{2M}$) should be handled with care. We treat these regions differently when dealing with $\epsilon_1$ and $\epsilon_2$. For $\epsilon_1$ (corresponding to spherical coordinate $\phi$) we handle the boundaries by linearly interpolating $\vect{v}[0]$ and $\vect{v}[M-1]$, as $\epsilon_1 = 0$ and $1$ point to identical directions. For $\epsilon_2$, on the other hand, we use nearest neighbor to handle the boundaries. In both cases, the formulation in Eq.~\ref{eq:normalization_linear} remains valid and thus we use the same architecture, as shown in Fig.~\ref{fig:network}.

Sampling from the learned distribution is again performed using inverse CDF sampling, as shown in Fig.~\ref{fig:interpolation} (bottom-right). The main difference is that the CDF is piecewise quadratic (instead of linear). Refer to the supplementary materials for the full derivation.

\vspace{-0.05in}
\paragraph*{Discussion} Our representation has similarities to action space discretization in reinforcement learning, as proposed by Tang and Agrawal~\shortcite{Tang_AAAI_2020}. However, there are two key differences. First, this technique discretizes the PDF, so evaluation and sampling are done at discrete locations. In contrast, we estimate PDF at discrete locations, but evaluation and sampling remain continuous. The second, more subtle difference is that they assume different dimensions of the action space are independent, representing the joint PDF as a product of marginal distributions. In our case, the dimensions are correlated, so our second PDF is conditional (see Eq.~\ref{eq:product_rule}).

\subsubsection{Connection to NIS} Our method addresses the problem of learning a PDF and sampling from it in a fundamentally different way than NIS~\cite{Muller_TOG_2019}. We focus on directly modeling a multidimensional PDF by estimating a series of marginal and conditional distributions; for example, in our approach, we estimate four 1D PDFs to model a four-dimensional PDF. Sampling is a byproduct of our system, performed through inverse CDF sampling. In contrast, NIS uses normalizing flows, where the primary process maps samples from a known (e.g., uniform) distribution to the target distribution through a set of coupling layers, with PDF evaluation being a byproduct of this sampling process. Specifically, NIS splits the input dimensions into two equally sized partitions, using one as input to the network to estimate the mapping function that warps the other dimensions. This process is repeated in the remaining coupling layers in an alternating pattern. 

Our factorization is motivated by the product relationship between joint and 1D PDFs, whereas NIS splits the input dimensions due to its use of normalizing flows with coupling layers. This fundamental difference in approach leads to a significant performance gap, even in the more constrained path guiding scenario with only two input dimensions. Specifically, the key difference in two dimensions is that \emph{evaluating} the PDF of a given sample in NIS requires evaluating the two networks sequentially, as the warped sample serves as the condition for the network in the next coupling layer. In contrast, we evaluate the PDF by executing the two networks in parallel. This formulation results in simpler gradients (avoiding backpropagation through two networks), lower training times, and faster inference, without sacrificing expressiveness (see supplementary material).

\vspace{-0.05in}
\subsection{Optimization with Radiance Caching}
\label{ssec:optimization}

Since our goal is to approximate the target distribution using our learned PDF, we optimize our networks by minimizing the KL divergence, defined as:

\vspace{-0.125in}
\begin{equation}
\label{eq:kl_div}
    D\big(p(\diri)\Vert \hat{p}_\netparam(\diri)\big) = \int_\Omega p(\diri) \log \frac{p(\diri)}{\hat{p}_\netparam(\diri)} \text{d}\diri
\end{equation}
\vspace{-0.125in}

\noindent where $p$ and $p_\netparam$ are the target and guiding distributions, respectively. Note that the guiding distribution is obtained through the product of the two 1D PDFs in Eq.~\ref{eq:product_rule}, evaluated using the two networks. Moreover, all the distributions are conditioned on both $\surface$ and $\diro$, but we remove the condition for simplicity of notation.

Minimizing this loss necessitates taking its derivative with respect to the parameters of the guiding distribution (the two networks parameters $\netparam = \{\netparamf, \netparams \}$) as follows:

\vspace{-0.15in}
\begin{equation}
    \nabla_\netparam D\big(p(\diri)\Vert \hat{p}_\netparam(\diri)\big) =  - \int_\Omega p(\diri) \nabla_\netparam \log \hat{p}_\netparam(\diri) \text{d}\diri,
\end{equation}
\vspace{-0.15in}

\noindent where we omit the gradient with respect to $\log p(\diri)$, since it is independent of $\netparam$. Calculating this derivative requires evaluating the integral which does not have a closed form solution. Therefore, similar to the existing methods~\cite{Dong_2023,Muller_TOG_2019}, we approximate it through MC integration as follows:

\vspace{-0.15in}
\begin{equation}
\label{eq:gradient_estimate}
    \langle \nabla_\netparam D\big(p(\diri)\Vert q_\netparam(\diri)\big) \rangle = -  \frac{1}{N} \sum_{i = 1}^N \frac{p(\diri_i)}{q(\diri_i)} \nabla_\netparam \log \hat{p}_\netparam(\diri_i)
\end{equation}
\vspace{-0.15in}

\noindent where the samples $\diri_i$ are drawn from $q(\diri_i)$, which could be BSDF importance sampling, guiding distribution, or a combination of the two. Moreover, the ideal target distribution is:

\vspace{-0.15in}
\begin{equation}
\label{eq:target_dist_sample}
    p(\diri_i) = \underbrace{\bsdfsym(\surface, \diro, \diri_i) L_i(\surface, \diri_i) |\cos(\theta_i)|}_{\text{integrand}} \underbrace{L_r(\surface, \diro)^{-1}}_{\text{normalization factor}},
\end{equation}
\vspace{-0.15in}

\noindent where the left term is the integrand in Eq.~\ref{eq:rendering_eq} and the normalization factor is to ensure the PDF integrates to one. 

Since the true incoming radiance $L_i(\surface, \diri_i)$ is not available, existing methods~\cite{Dong_2023,Muller_TOG_2019} estimate it using MC samples. Unfortunately, this increases the variance of the gradients, and consequently makes the optimization difficult, particularly in complex scenes. Additionally, the normalization term is the reflected radiance, a quantity that we are aiming to estimate. Current techniques ignore this factor with the hope that the scale is canceled when using optimization approaches, such as Adam ~\cite{Kingma_adam_2015}, that work with the ratio of the current and historic gradients. This is true if the optimization is performed only at a fixed $\surface$ and $\diro$, as the normalization factor will be the same for all the gradients. However, in practice, samples from a large number of surface locations with various outgoing directions are used to compute the gradients. Therefore, ignoring the normalization factor could potentially hurt the performance of the optimization.

\begin{figure}[t]
\centering
\includegraphics[width=1.0\linewidth, scale=1.0, angle=-0]{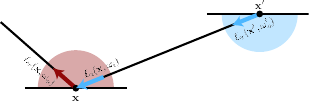}
\vspace{-0.25in}
\caption{Computing the target distribution requires obtaining $L_r(\surface, \diro)$ and $L_i(\surface, \diri_i)$ (see Eq.~\ref{eq:target_dist_sample}). We use a neural network that takes location and direction as the input and estimates the cached reflected radiance along that particular ray. By evaluating the network at the current ($\surface^\prime, \diro^\prime$) and next ($\surface, \diri_i$) intersection points, we obtain an estimate of the reflected $L_r$ and incoming $L_i$ radiance. Note that here the incoming radiance at $\surface, \diri_i$ is equal to the reflected radiance at $\surface^\prime, \diro^\prime$. }
\vspace{-0.3in}
\label{fig:radiance_caching}
\end{figure}

To address the first issue and reduce the variance of the gradient, we propose to utilize radiance caching, a well-explored topic in rendering. In particular, we follow the technique by M\"{u}ller et al.~\shortcite{Muller_TOG_2021} (NRC) and cache the radiance using a small MLP network that takes the surface location and ray direction as the input and estimates the corresponding radiance. Similar to NRC, instead of caching the incoming radiance at the current location $L_i(\surface, \diri_i)$, we learn the reflected radiance at the next intersection point $L_r(\surface^\prime, \diro^\prime)$. Despite the equivalence of these two quantities (refer to Fig.~\ref{fig:radiance_caching}), we observe superior performance in learning $L_r$ over $L_i$ due to differences in input coordinates. In our system, the network $f_\nrcparam$ takes the surface location at the next intersection point along with the corresponding outgoing direction and estimates the reflected radiance, i.e., $\hat{L}_r (\surface^\prime, \diro^\prime) = f_\nrcparam (\surface^\prime, \diro^\prime)$. Since $L_i(\surface, \diri_i) = L_r(\surface^\prime, \diro^\prime)$, we can leverage this network to estimate the incoming radiance in Eq.~\ref{eq:target_dist_sample}.

Since the network estimates the reflected radiance, it can be used to estimate the normalization factor in Eq.~\ref{eq:target_dist_sample}, thereby addressing the second problem. To summarize, using our radiance caching network, the target distribution at $\diri_i$ is estimated as follows:

\vspace{-0.2in}
\begin{equation}
\label{eq:est_target_dist_sample}
    \hat{p}(\diri_i) = \bsdfsym(\surface, \diro, \diri_i) f_\nrcparam(\surface^\prime, \diro^\prime) |\cos(\theta_i)| \quad f_\nrcparam(\surface, \diro)^{-1}.
\end{equation}
\vspace{-0.2in}

We use this estimated target distribution $\hat{p}(\diri_i)$ in place of $p(\diri_i)$ in Eq.~\ref{eq:gradient_estimate} to estimate the gradient of the KL divergence in training.

Note that theoretically, caching radiance and using it as $L_i$ should not help with variance reduction; the variance will just be passed from here to radiance caching optimization. However, as shown in Fig.~\ref{fig:radiance_caching_ablation}, we find this to be helpful in practice. This could be attributed to NRC's~\cite{Muller_TOG_2021} highly effective approach to deal with the variance through the relative $\mathcal{L}^{2}$ loss, as well as the strategy to learn $L_r$ as opposed to $L_i$.

\begin{table*}[ht]
\setlength\tabcolsep{3pt} % default value: 6pt
\centering
\caption{Equal-time comparisons measured in relMSE (lower is better) for \rev{seven} representative scenes. We compare our approach with nearest neighbor (DF-N) and linear (DF-L) interpolation against unidirectional path tracing (PT), PPG~\cite{Muller_CGF_2017}, Variance~\cite{Rath_TOG_2020}, NIS~\cite{Muller_TOG_2019}, and NPM~\cite{Dong_2023}. We color code the \colorbox{red!40}{first}, \colorbox{orange!40}{second}, and \colorbox{yellow!40}{third} lowest numbers.}
\vspace{-0.05in}
\footnotesize
\label{tab:equal_time}
\begin{tabular}{lcccccccccccccc}
\toprule
 & \multicolumn{2}{c}{PT} & \multicolumn{2}{c}{PPG} & \multicolumn{2}{c}{Variance} & \multicolumn{2}{c}{NIS} & \multicolumn{2}{c}{NPM} & \multicolumn{2}{c}{DF-N} & \multicolumn{2}{c}{DF-L} \\ 
\midrule
\textsc{Bathroom} & 1.1140 & 3774spp & 0.5152 & 1598spp & 0.5729 & 1445spp & \cellcolor{yellow!40}0.2738 & 789spp & 0.3761 & 1507spp & \cellcolor{orange!40}0.2601 & 1166spp & \cellcolor{red!40}0.2106 & 1131spp \\
\midrule
\textsc{Bedroom} & 0.0270 & 4759spp & 0.0389 & 1634spp & 0.0318 & 1591spp & 0.0201 & 928spp & \cellcolor{yellow!40}0.0182 & 1830spp & \cellcolor{orange!40}0.0167 & 1365spp & \cellcolor{red!40}0.0138 & 1328spp \\
\midrule
\textsc{Breakfast} & 0.1668 & 6675spp & 0.1578 & 1978spp & 0.1080 & 1981spp & \cellcolor{yellow!40}0.0263 & 1414spp & 0.0286 & 3267spp & \cellcolor{orange!40}0.0199 & 2161spp & \cellcolor{red!40}0.0171 & 2046spp \\
\midrule
\textsc{Kitchen} & 0.0821 & 4697spp & 0.0386 & 2644spp & 0.0323 & 2240spp & 0.0663 & 960spp & \cellcolor{yellow!40}0.0196 & 2091spp & \cellcolor{orange!40}0.0186 & 1427spp & \cellcolor{red!40}0.0146 & 1382spp \\
\midrule
\textsc{Salle de Bain} & 0.0654 & 4969spp & 0.1067 & 1462spp & 0.0625 & 1469spp & 0.0251 & 937spp & \cellcolor{yellow!40}0.0233 & 2249spp & \cellcolor{orange!40}0.0193 & 1384spp & \cellcolor{red!40}0.0155 & 1355spp \\
\midrule
\textsc{Staircase} & 2.5989 & 5403spp & \cellcolor{yellow!40}0.6820 & 2584spp & 1.1779 & 2510spp & 1.3275 & 1202spp & 1.1155 & 1981spp & \cellcolor{orange!40}0.3494 & 1763spp & \cellcolor{red!40}0.2861 & 1729spp \\
\midrule
\textsc{Veach Door} & 14.182 & 6054spp & 3.0940 & 1755spp & 9.3482 & 1738spp & 1.9415 & 1116spp & \cellcolor{yellow!40}1.8092 & 2828spp & \cellcolor{orange!40}0.7132 & 1524spp & \cellcolor{red!40}0.4129 & 1454spp \\
\bottomrule
\end{tabular}
\vspace{-0.1in}
\end{table*}

\vspace{-0.015 in}
\paragraph*{Discussion} We note an interesting connection between our method and actor-critic techniques in reinforcement learning~\cite{Sutton_Book_2018,Barto_TSMC_1983}. In reinforcement learning, the goal is to minimize the expected reward over all the trajectories. During training, this is often approximated using sampled trajectories making the gradients highly noisy. The idea behind actor-critic methods is to cache the reward (critic) to reduce the variance of gradient for optimizing the policy (actor). In this context, our radiance caching network $f_\nrcparam$ is the critic, while the distribution learning network $f_\netparam$ can be thought of as the actor. 

A similar caching strategy has been applied to path guiding by Dahm and Keller~\shortcite{Dahm_2017} using $Q$-learning, which is a closely related, but different concept. Pantaleoni~\shortcite{Pantaleoni_2020} further explores caching by comparing different basis functions and data structures on both unbiased and biased alternatives. The main discerning characteristics of our approach are: 1) modeling the cache implicitly, which avoids spatial discretization artifacts, and 2) caching the outgoing radiance $L_r$ with $f_\nrcparam$ to more effectively learn the product of incoming radiance $L_i$ and BSDF with $f_\netparam$.

\begin{figure*}
\centering
\includegraphics[width=\textwidth]{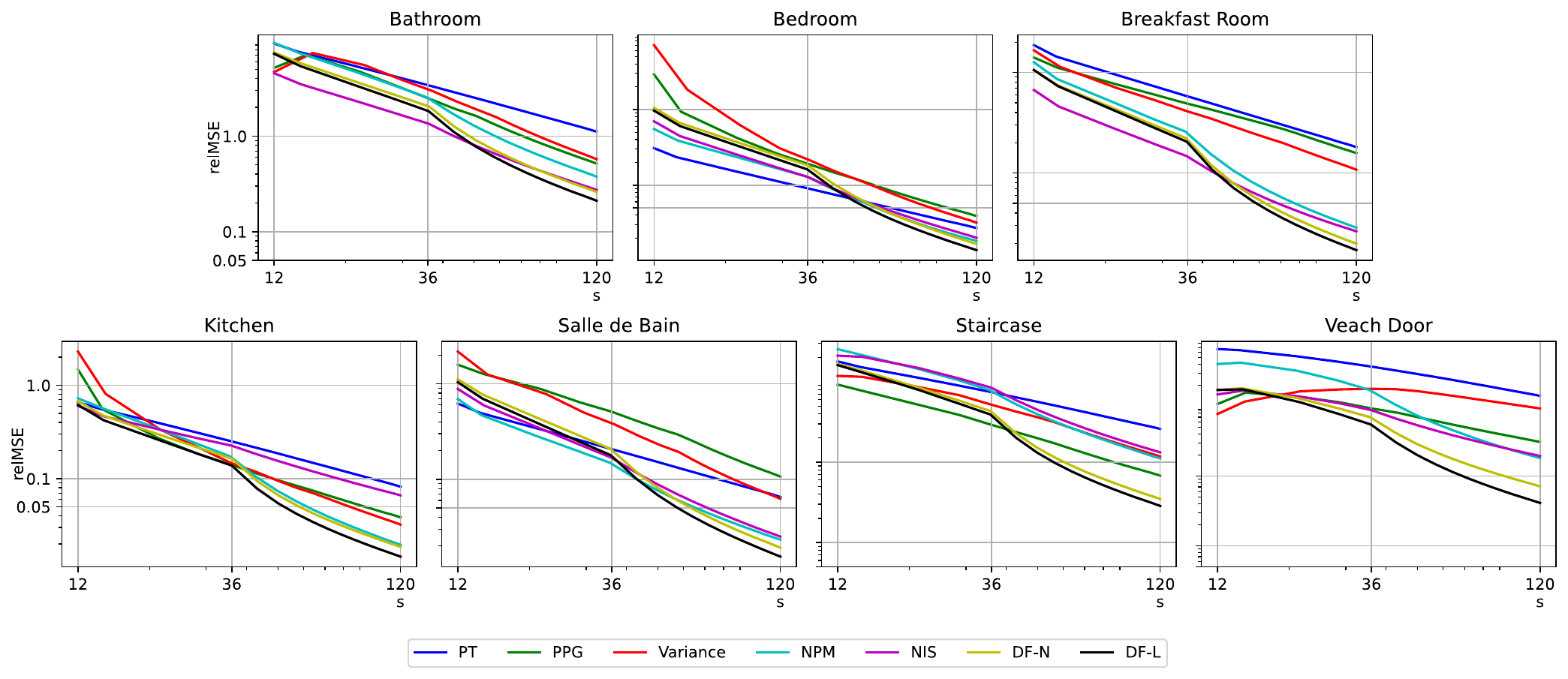}
\vspace{-0.2in}
\caption{We show convergence plot of all the approaches on the \rev{seven} scenes from 12 to 120 seconds (s). The 36 s line is where all the approaches (except the unidirectional path tracer) stop learning and use the learned distributions to sample the remaining paths. }
\vspace{-0.2in}
\label{fig:timeconvergenceplots}
\end{figure*}
\vspace{-0.05in}
\section{Implementation}

\paragraph*{Online Optimization with Radiance Caching}
We optimize our approach in an online fashion using the prototype GPU-accelerated renderer of~\cite{Dong_2023}. For all experiments, we simultaneously train and guide paths for the first 30\% of the allocated budget of time or spp; after which optimization is halted for the remainder of the task. To increase efficiency, we reuse rendering rays for training our radiance caching network at the end of each rendered frame. The training data is split into at least four mini batches of equal size to ensure an aggressive optimization of $f_\nrcparam$ before it is used to train $f_{\netparamf}$ and $f_{\netparams}$. We use Adam~\cite{Kingma_adam_2015} with default parameters and a fixed learning rate of $10^{-2}$ for $f_\nrcparam$. For the PDF estimation networks, we found that increasing the learning rate further to $3\times10^{-2}$ is beneficial and stable thanks to our smoother cached objectives. We leverage the learned distributions to guide 70\% of paths using our method, while the remainder of the paths is generated with unbiased BSDF sampling to allow exploration. We provide a memory and runtime analysis of our approach in the supplementary.

\paragraph*{Networks}
Our networks, implemented in tiny-cuda-nn~\cite{tiny-cuda-nn}, share the same underlying architecture of 3 hidden layers of 64 neurons and ReLU intermediary activation functions. We encode the inputs using ~\cite{Dong_2023}'s learnable dense grid encoding for the intersection position, and spherical harmonics with degree 4 for normalized outgoing ray direction $\diro$. Similar to existing methods, we additionally provide normal and roughness as the input to the networks and encode them through one-blob encoding~\cite{Muller_TOG_2019} using 4 bins. Our conditional PDF estimation network $f_{\netparams}$ requires the extra input of $\epsilon_1$ for which we use triangle wave encoding~\cite{Muller_TOG_2021} with 12 frequencies. We set the discretization resolution $M_1$ to 32 and $M_2$ to 16. The reduction in resolution in the second dimension is due to the smaller angular range of $\epsilon_2$ corresponding to $\theta$ (0 to $\pi$). 
\vspace{-0.05in}
\section{Results}
\label{sec:results}
In this section, we first describe the evaluation scenes, then show extensive visual and numerical comparisons to demonstrate the effectiveness of our approach. For all the results, we use a unidirectional path tracer with BSDF sampling and disable next-event estimation (NEE), as done in prior methods~\cite{Muller_TOG_2019,Zhu_TOG_2021,Dong_2023}, to better illustrate the effectiveness of path guiding. Additionally, we disable Russian roulette and limit the path length to 6. We perform all comparisons and ablations on an AMD Ryzen 5800X CPU and an NVIDIA RTX3080 GPU. Here, we report quantitative results using the trimmed relative mean squared error~\cite{rousselle2011adaptive} (relMSE) with a threshold of 0.1\% averaged over 10 independent runs. Evaluation using additional metrics can be found in the supplementary.

\vspace{-0.1in}
\subsection{Scenes}

To better showcase complex lighting scenarios that are commonplace in real-world applications, \rev{seven} representative scenes are used throughout this section. Here, we briefly describe each scene's lighting setup and challenges. The \textsc{Bathroom} and \textsc{Salle de Bain} scenes show examples of bathrooms featuring cone-shaped and cube-shaped light fixtures, respectively. In order to reach a light source, paths require finer guidance from the learnable methods given the spacing between lights and their directional component. The semi-transparent curtains of the \textsc{Bedroom} scene create a barrier to the light sources, resulting in indirect illumination for most of the paths, leaving only a small opening at the sides for direct lighting. The \textsc{Breakfast} scene features directional lighting through the blinds in addition to two high-intensity light fixtures that are focused on the table. When combined with the metallic BSDFs of the teapot and cups, it results in a small quantity of viable paths to light, increasing complexity. Similarly, the \textsc{Kitchen} scene also features a directional light coming from the window in addition to a second light source on top of the microwave. The \textsc{Staircase} is illuminated by three small high-intensity light fixtures placed at an angle that result in overlapping conic sections on the main atrium. Lastly, the \textsc{Veach Door} scene showcases an example of indirect lighting, as the scene is lit through the reflections on the right wall. Three separate light sources sit behind the partially-open door.

\vspace{-0.05in}
\subsection{Comparisons}
\label{sec:comparisons}

Throughout this section, we compare our approach against several state-of-the-art path guiding methods. Specifically, we show comparisons against M\"{u}ller et al.~\shortcite{Muller_CGF_2017} (PPG), Rath et al.~\shortcite{Rath_TOG_2020} (Variance), M\"{u}ller et al.~\shortcite{Muller_TOG_2019} (NIS), and Dong et al.~\shortcite{Dong_2023} (NPM). We implement all methods on the fast prototype GPU path tracer provided by Dong et al.~\shortcite{Dong_2023}. We show the results of two variants of our distribution factorization approach with nearest (DF-N) and linear (DF-L) interpolation. We implement NIS with two fully-fused networks ($L=2$) using piecewise quadratic coupling layers and fixed \rev{bin size and} resolutions of $32\times 16$ to match the capacity of our approach. Note that NIS, NPM and our method are product-based (learn the product of radiance and BSDF), but the other two methods are radiance-based.

\paragraph*{Equal Time} We begin by showing equal-time numerical and visual comparisons against the other approaches in Table~\ref{tab:equal_time} and Fig.~\ref{fig:equaltime} (see supplementary material for equal sample comparisons). To obtain these results, we provide a budget of 120 seconds to all the approaches, including the standard path tracer (PT). As shown, our method with linear interpolation produces the best results in all the scenes, with the nearest-neighbor variant being second-place. Faster methods such as PPG, Variance and NPM can render significantly more samples than our approach in some scenes, but fail to model PDFs accurately. In contrast, NIS traces fewer samples than our method in the same time.

\begin{figure*}
\centering
\includegraphics[width=0.87\textwidth]{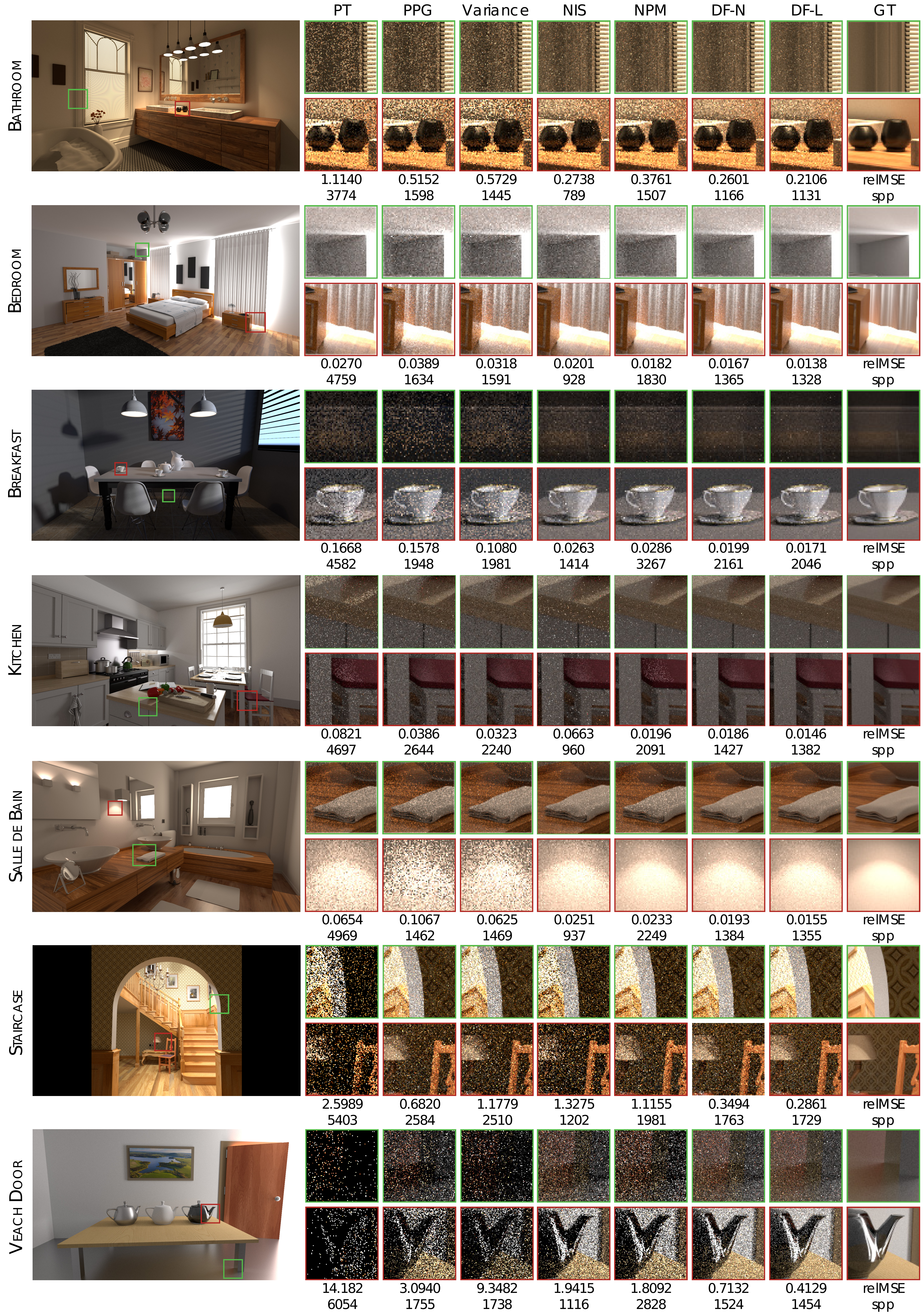}
\vspace{-0.10in}
\caption{Equal-time comparison against M\"{u}ller et al.~\shortcite{Muller_CGF_2017} (PPG), Rath et al.~\shortcite{Rath_TOG_2020} (Variance), M\"{u}ller et al.~\shortcite{Muller_TOG_2019} (NIS), Dong et al.~\shortcite{Dong_2023} (NPM), and our method with nearest neighbor (DF-N) and linear (DF-L) interpolation. The time budget for all the methods is 120 seconds, with a training budget of 30\%.}
\label{fig:equaltime}
\end{figure*}

\begin{figure}
\centering
\includegraphics[width=1.0\linewidth, scale=1.0, angle=-0]{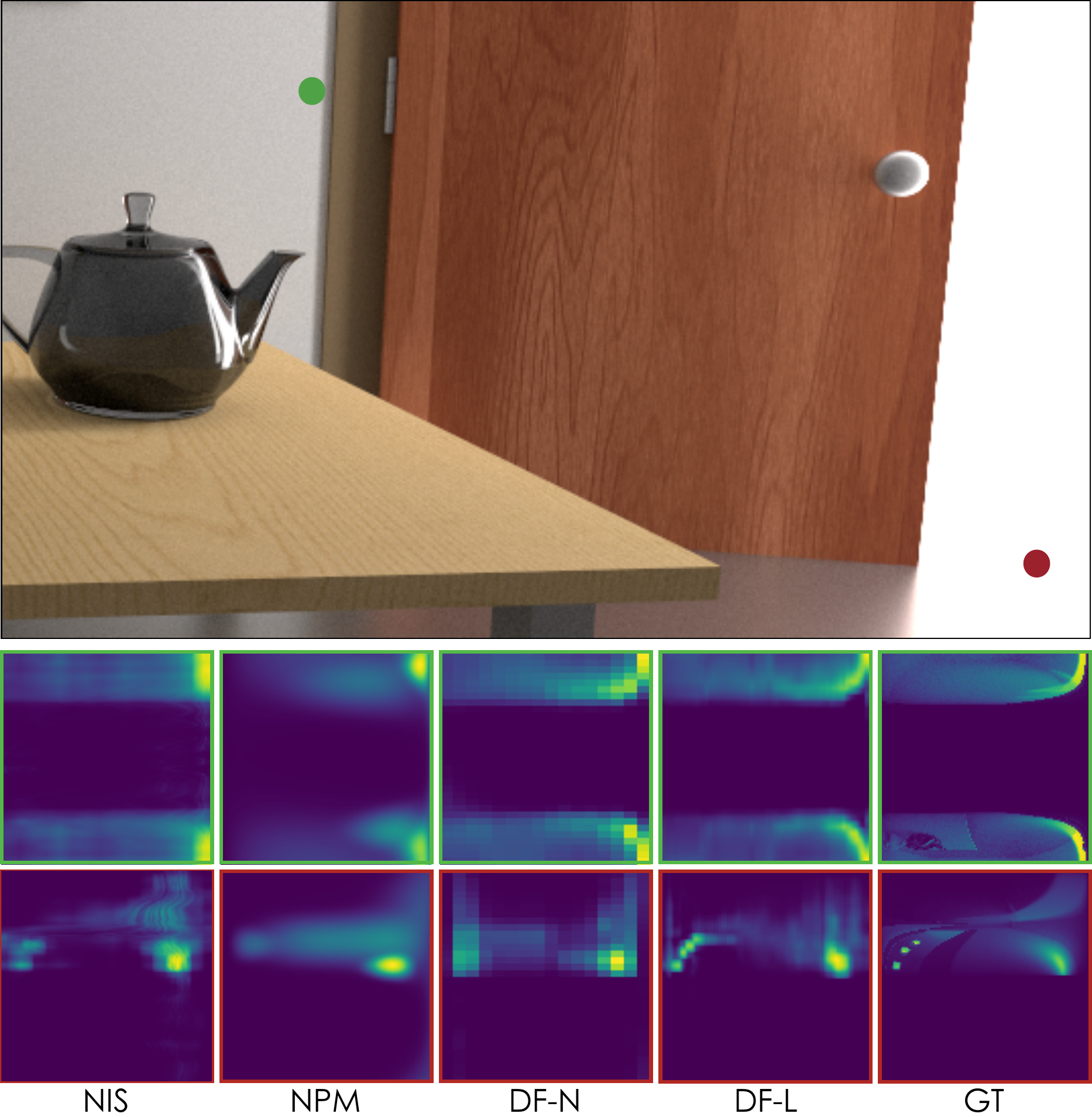}
\vspace{-0.2in}
\caption{Equal-time comparison of learned distributions for the \textsc{Veach Door} scene. The time budget for all methods is 120s.}
\vspace{-0.25in}
\label{fig:learnedvis}
\end{figure}

We further demonstrate the encoding quality of our approach in an equal-time setting by comparing the learned PDFs of our method against NIS~\cite{Muller_TOG_2019} and NPM~\cite{Dong_2023} on the \textsc{Veach Door} scene in Fig.~\ref{fig:learnedvis}. NIS trains on fewer noisy MC samples, which results in a less detailed representation. NPM produces blurry estimations, leading to less samples reaching the light. Our approach with nearest-neighbor interpolation (DF-N) roughly approximates the target distribution, but fails to encode the light sources with detail due to a resolution limitation. In contrast, DF-L produces sharper approximations that correlate the most with the reference. Note that we do not show visualizations for M\"{u}ller et al.~\shortcite{Muller_CGF_2017} or Rath et al.~\shortcite{Rath_TOG_2020} since these methods encode the radiance while our approach is designed for the product with BSDF.

\vspace{-0.05in}
\paragraph*{Convergence Analysis} To understand the convergence behavior of our approach, we plot the relMSE of the rendered images in equal-time (Fig.~\ref{fig:timeconvergenceplots}) and equal-sample (supplementary) settings. Overall, our approach produces better results than all the other methods in all cases. We note that, in some scenes (e.g. \textsc{Salle de Bain}), other methods are able to learn a better map initially, but our results improve significantly once the training phase ends and we can trace more samples. Moreover, our method with linear interpolation is consistently better than nearest neighbor. \rev{Finally, the improvement becomes larger over time, as we learn better guiding distributions than the other approaches.}

\vspace{-0.1in}
\subsection{Ablations}

\begin{figure}
\centering
\includegraphics[width=0.95\linewidth, scale=1.0, angle=-0]{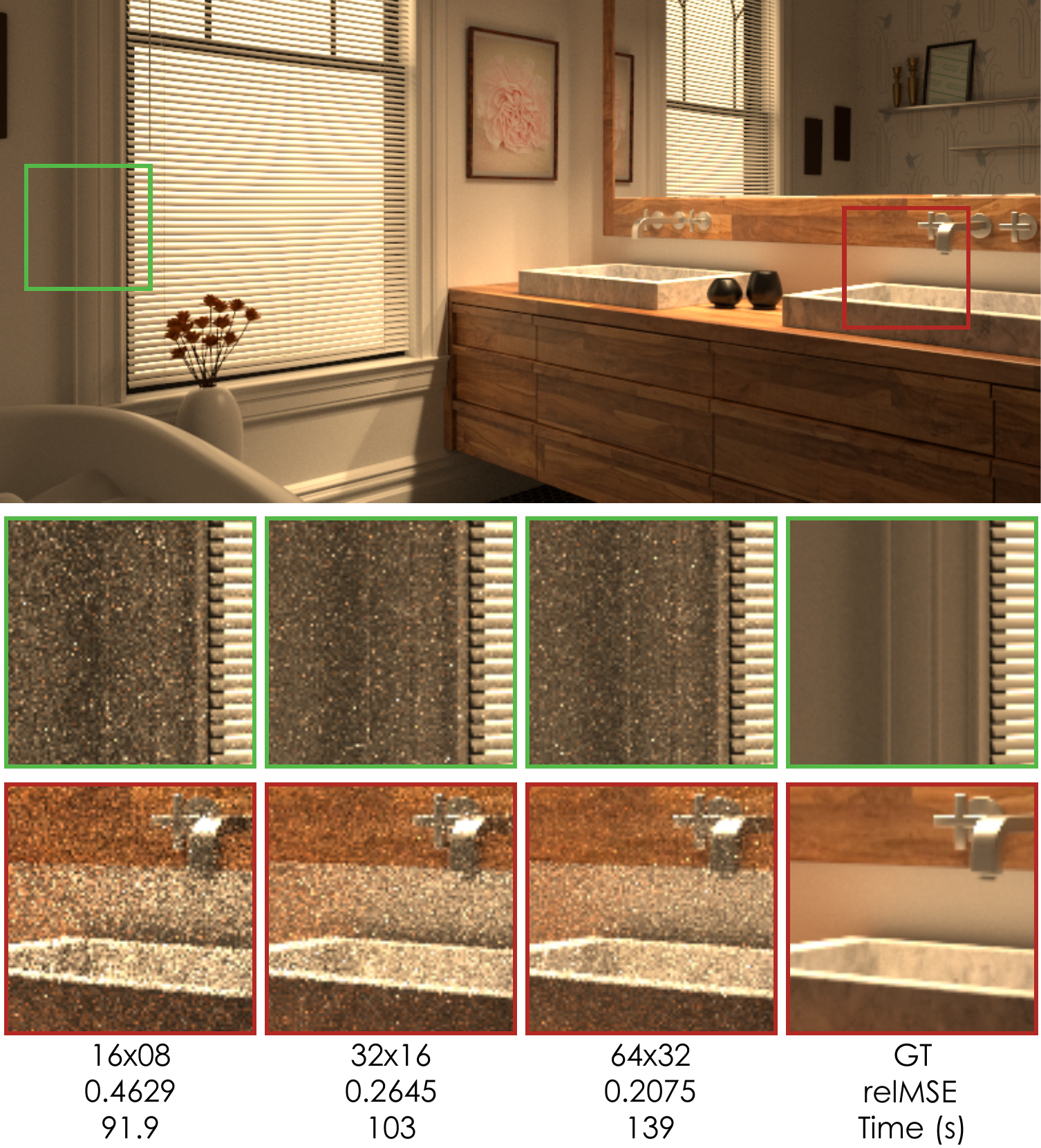}
\vspace{-0.05in}
\caption{Effect of discretization resolution $M_1\times M_2$ on the quality of the results. As the resolution increases, the quality improves, but this comes at the cost of increased computational complexity.}
\label{fig:discretization_resolution}
\end{figure}

\paragraph*{Discretization Resolution} Here, we evaluate the effect of resolution ($M_1$ and $M_2$) on the quality of results. We show our rendered results with 750 spp using three different resolutions in Fig.~\ref{fig:discretization_resolution}. Note that $M_2$ is always half of $M_1$ as it corresponds to $\theta$ which is between 0 and $\pi$. As seen, by increasing the resolution, our networks can model the target distribution more accurately, leading to better results. Particularly, increasing the resolution from $16\times 8$ to $32\times 16$ significantly improves the results, but the improvement beyond this resolution is smaller. Given the computational cost associated with higher resolutions, we choose $32\times 16$ to generate all the results.

\paragraph*{Optimization with Radiance Caching} We further evaluate the impact of caching radiance during optimization in Fig.~\ref{fig:radiance_caching_ablation} for the \textsc{Veach Door} scene (and for all scenes in the supplementary). The impact of cache varies by scene, with higher gains in scenes with complex light transport. For the \textsc{Veach Door}, using the cached radiance to estimate both $L_r$ and $L_i$ in the target distribution (Eq.~\ref{eq:est_target_dist_sample}) significantly improves the results.

\vspace{-0.05in}
\section{Conclusion, Limitations, and Future Work}

In this paper, we have presented a novel neural path guiding method. We first factorize the distribution over the 2D directional domain into two 1D PDFs and estimate them using two separate neural networks at discrete locations. The PDF at an arbitrary coordinate is then obtained through interpolation. We explore two interpolation strategies and discuss evaluation and sampling process for each. Moreover, we propose a strategy to reduce the variance and normalize the samples used as the target distribution during optimization. Specifically, we demonstrate that we can cache the radiance using an additional network and use the estimates for incoming radiance and normalization. We demonstrate the effectiveness of our approach with extensive experiments.

\begin{figure}
\centering
\includegraphics[width=0.95\linewidth, scale=1.0, angle=-0]{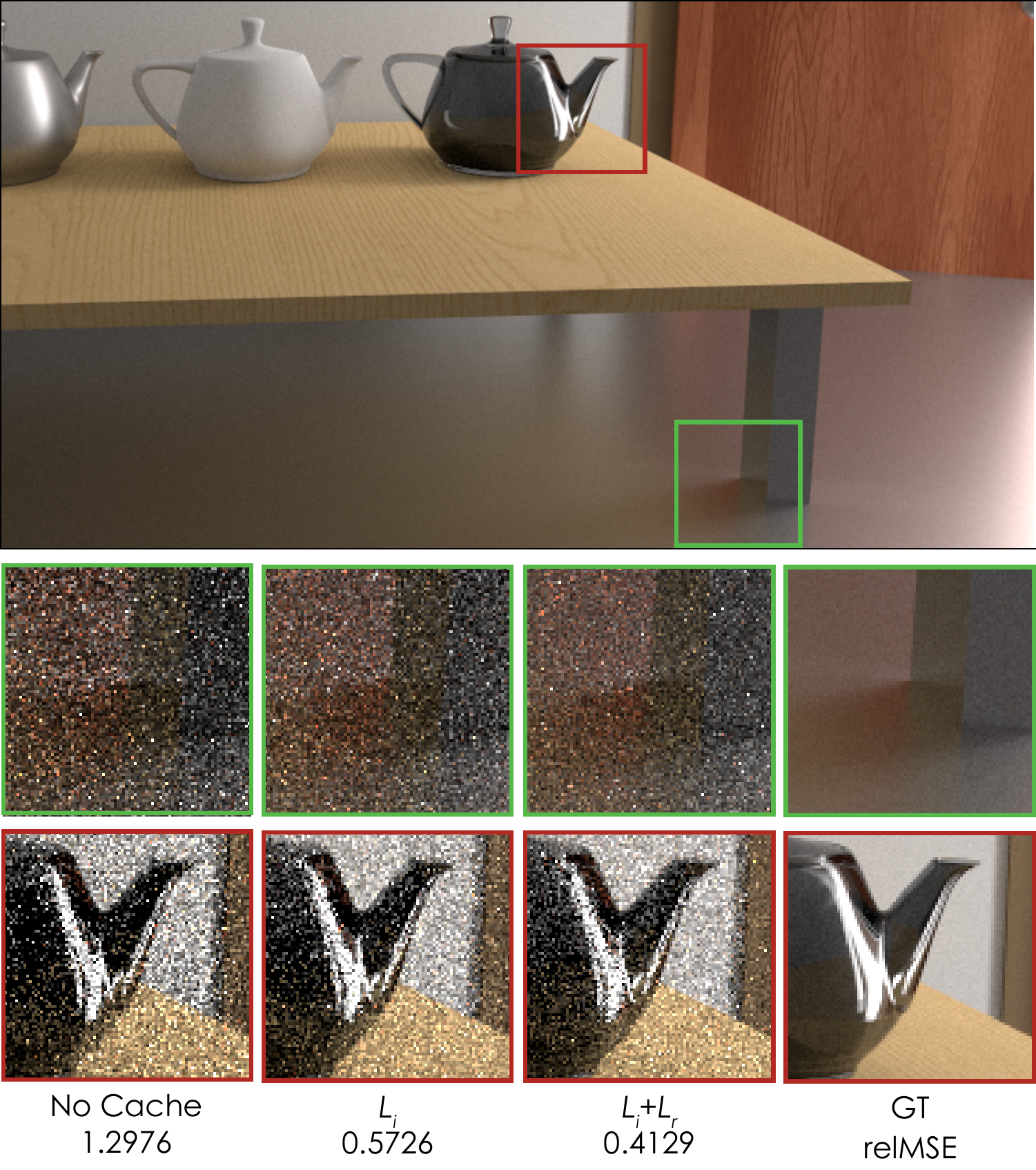}
\vspace{-0.05in}
\caption{Effect of radiance caching during equal-time comparison of 120s on the \textsc{Veach Door}. We compare the results of our approach ($L_r + L_i$) against not using the cache (similar to existing methods) as well as using the cache only for $L_i$ in Eq.~\ref{eq:est_target_dist_sample}. Using the cache for both $L_i$ and $L_r$ improves the results.}
\label{fig:radiance_caching_ablation}
\end{figure}

\begin{figure*}
\centering
\includegraphics[width=\linewidth, scale=1.0, angle=-0]{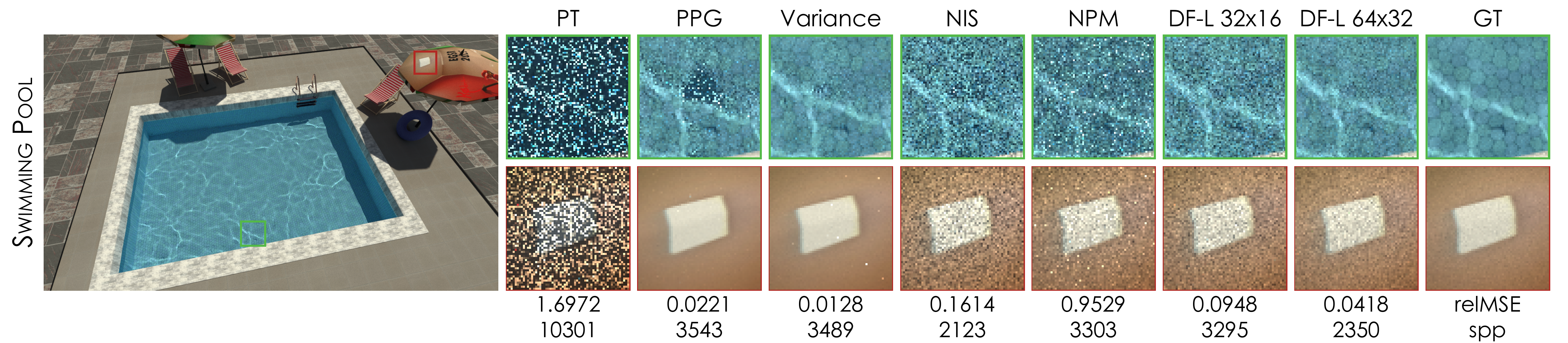}
\vspace{-0.2in}
\caption{\rev{Equal-time comparison (120s) on the \textsc{Swimming Pool} scene. The pool shows direct lighting from an environment map that models the Sun forming caustics in an outdoor swimming pool. Our approach struggles in this case as it does not sharply encode the directional light from the Sun. This is because of our fixed discretization resolution since, as shown, results improve as we increase the resolution from 32$\times$16 to 64$\times$32.}}
\vspace{-0.2in}
\label{fig:limitation-pool}
\end{figure*}

\rev{While our approach works well in most tested scenarios, it uses a fixed resolution, and thus struggles to encode features that are significantly smaller than each bin. We visualize this limitation on an equal-time comparison of the outdoor \textsc{Swimming Pool} scene in Fig.~\ref{fig:limitation-pool}. As shown, our method does not properly model the Sun, producing results with excessive noise. This problem is directly related to the resolution since increasing it from 32$\times$16 to 64$\times$32 improves the results. Similarly, NIS~\cite{Muller_TOG_2019} (with fixed bin size) suffers from the same limitation, but produces slightly worse results because of higher computational cost. In contrast, PPG~\cite{Muller_CGF_2017} and Variance~\cite{Rath_TOG_2020} quickly adapt their data structures to focus on the directional light of the Sun, resulting in the least noise. Although NPM~\cite{Dong_2023} uses a continuous representation, surprisingly, it struggles the most here, as its optimization becomes unstable with the concentrated high intensity illumination. In the future, it would be interesting to combine our approach with explicit spatial data structures to vary the resolutions based on the complexity. Alternatively, we could adopt the variable bin size strategy, as proposed by  M\"{u}ller et al.~\shortcite{Muller_TOG_2019}, to increase the flexibility of fixed resolution.}

\begin{figure}
\centering
\includegraphics[width=0.95\linewidth, scale=1.0, angle=-0]{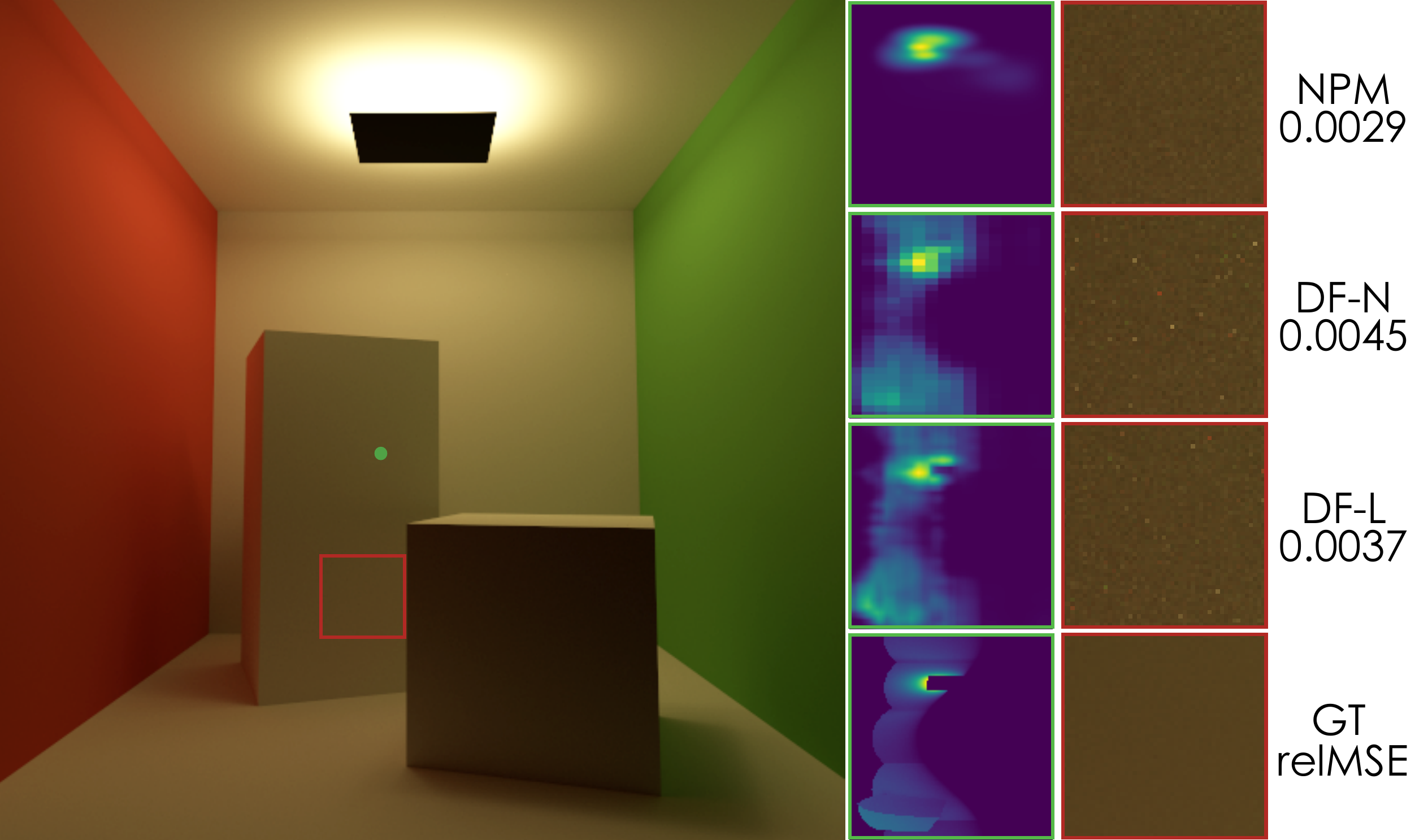}
\vspace{-0.05in}
\caption{Equal-time comparison (30 seconds budget) on the \textsc{Cornell Box} with flipped light. Although DF-N and DF-L encode higher quality PDFs, Dong et al.~\cite{Dong_2023} (NPM) generate the result with least noise because they trace more samples.}
\vspace{-0.25in}
\label{fig:limitation}
\end{figure}

\rev{In addition, we designed our approach to be highly effective specifically when handling complex scenes. On simple scenes with easy to model light transport, the advantage of our method is not significant.} For such cases, when using a fast GPU path tracer, simpler and faster methods might produce better results because of their ability to trace more samples. Fig.~\ref{fig:limitation} demonstrates this phenomenon through an equal-time experiment of 30 seconds using the \textsc{Cornell Box} with flipped light. \rev{Although} the learned PDF of the two variants of our approach better correlate with the target distribution compared to Dong et al.~\shortcite{Dong_2023} (NPM), NPM's ability to throw more samples results in less noise overall.

\vspace{-0.1in}
\section*{Acknowledgments}
This project was funded in part by the NSF CAREER Award $\#2238193$. We are grateful to Dong et al.~\shortcite{Dong_2023} for releasing the source code of their work. We would like to thank the following artists for sharing their scenes and models that appear in our figures: Mareck (\textsc{Bathroom}), SlykDrako (\textsc{Bedroom}), Wig42 (\textsc{Breakfast}, \textsc{Staircase}), Jay-Artist (\textsc{Kitchen}), nacimus (\textsc{Salle de Bain}), Benedikt Bitterli (\textsc{Veach Door}, \textsc{Cornell Box}) and CN Entertainment, LordSamueliSolo (\textsc{Swimming Pool}).

\vspace{-0.1in}
\bibliographystyle{eg-alpha-doi} 
\bibliography{bibliography}   

\end{document}